\documentstyle[12pt]{article}

\def\be{\begin{equation}}
\def\ee{\end{equation}}
\begin{document}

\begin{titlepage}
\hskip 11cm \vbox{ \hbox{DESY 11-191}  }
\vskip 3cm

\centerline{\large \bf
BFKL equation for the adjoint representation of the gauge group}
\centerline{\large \bf in the next-to-leading approximation
at $N=4$ SUSY $^{\dag}$}

\date{}

\begin{center}
V.S. Fadin {$^{*}$, L.N. Lipatov $^{**}$
\\
Universit\"{a}t Hamburg,\\
II. Institut f\"{u}r Theoretische Physik,\\
Luruper Chaussee, 149, D-22761 Hamburg,
\\
{*} Budker Nuclear Physics Institute\\
and Novosibirsk State University,\\
630090, Novosibirsk, Russia
\\
{**} Petersburg Nuclear Physics Institute\\
and St.Petersburg State University,\\
Gatchina, 188300, St.Petersburg, Russia}
\end{center}
\vskip 15.0pt
\centerline{\bf Abstract}

\noindent

We calculate the eigenvalues of the next-to-leading
kernel for the BFKL equation in the adjoint representation
of the gauge group $SU(N_c)$ in the N=4 supersymmetric Yang-Mills model.
These eigenvalues are used to obtain the high energy behavior of
the remainder function for the 6-point scattering
amplitude with the maximal helicity violation in the
kinematical regions containing the
Mandelstam cut contribution. The leading and next-to-leading singularities
of the corresponding collinear anomalous dimension are calculated
in all orders of perturbation theory. We compare our result with the
known collinear limit and
with the recently suggested ansatz for the remainder
function in three loops and obtain the full agreement
providing that the numerical parameters in this anzatz are chosen in
an appropriate way.


\noindent
\vskip 3cm
\vfill \hrule \vskip.3cm \noindent $^{\dag}${\it The work was supported
in part  by  grant  14.740.11.0082 of Federal Program
``Personnel of Innovational Russia,'' in part
by  RFBR grants 10-02-01238, 10-02-01338-a.}

\end{titlepage}
\section{Introduction}

In the Regge pole model  scattering amplitudes at large energies
$\sqrt{s} $ and fixed momentum transfers $\sqrt{-t}$ have the
form~\cite{Grib1} \be A_{Regge}^p(s,t)=\xi _p(t)\,s^{1+\omega
_p(t)}\,\gamma ^2(t)\,,\,\,\xi_p(t)=e^{-i\pi \omega_p (t)}-p\,, \ee
where $p=\pm 1$ is the signature of the reggeon with the trajectory
$\omega _p(t)$ and $\gamma ^2(t)$ represents the product of reggeon
vertices. The Pomeron is the Regge pole of the $t$-channel partial
wave $f_\omega (t)$ with vacuum quantum numbers and the positive
signature describing an approximately constant behaviour of total
cross-sections for the hadron-hadron scattering. S. Mandelstam
demonstrated, that the Regge poles generate  cut singularities in
the $\omega$-plane~\cite{Mand}.

In the leading logarithmic approximation (LLA) the scattering
amplitude at high energies in QCD has the Regge form~\cite{BFKL} \be
M_{AB}^{A^{\prime }B^{\prime }}(s,t)=M_{AB}^{A^{\prime }B^{\prime
}}(s,t)|_{Born}\,s^{\omega (t)}\,, \ee where $M_{Born}$ is the Born
amplitude and the gluon Regge trajectory is given below \be \omega
(-|q|^2)=-\frac{\alpha _s N_c}{4\pi ^2}\, \int
d^2k\,\frac{|q|^2}{|k|^2|q-k|^2}
 \approx
-\frac{\alpha _sN_c}{2\pi }\,\ln \frac{|q^2|}{\lambda ^2}\,.  \ee
Here $\lambda$ is the infrared cut-off.  In the multi-Regge
kinematics, where the pair energies $\sqrt{s_k}$ of the produced
gluons are large in comparison with momentum transfers $|q_i|$, the
production amplitudes in LLA are constructed from products of the
Regge factors $s_k^{\omega (t_k)}$ and effective
reggeon-reggeon-gluon vertices $C_\mu (q_r, q_{r+1})$~\cite{BFKL}.
The amplitudes satisfy the Steinmann relations and the $s$-channel
unitarity incorporated in bootstrap equations~\cite{Balitskii:1979}.

The knowledge of $M_{2\rightarrow 2+n}$ allows one to construct
the BFKL equation for the Pomeron
wave
function using analyticity, unitarity, renormalizability and crossing
symmetry~\cite{BFKL}.
The integral kernel of this equation has the property of the holomorphic
separability~\cite{int1}
and is invariant under the M\"obius
transformations~\cite{moeb}.
The generalization of this equation to a composite state of several
gluons~\cite{BKP} in the multi-color QCD
leads to an integrable
XXX model~\cite{int} having a duality symmetry~\cite{dual}.

The next-to-leading correction to the color singlet kernel in QCD is
also calculated~\cite{FL}. Its eigenvalue contains non-analytic
terms proportional to $\delta _{n,0}$ and $\delta_{n,2}$, where $n$
is the conformal spin of the M\"{o}bius group. But in the case of
the $N=4$ extended supersymmetric gauge model these Kronecker
symbols are canceled leading to an expression having the properties
of the hermitian separability~\cite{trajN4} and maximal
transcendentality~\cite{KL}. The last property allowed to calculate
the anomalous dimensions of twist-two operators up to three
loops~\cite{KLOV, KLOV1}. It turns out, that evolution equations for
the so-called quasi-partonic operators are integrable in $N=4$ SUSY
at the multi-color limit~\cite{integrQP}. The $N=4$ four-dimensional
conformal field theory according to the Maldacena guess is equivalent to
the superstring model living on the anti-de-Sitter 10-dimensional
space~\cite{Malda, GKP, W}. Therefore the Pomeron in N=4 SUSY is
equivalent to the reggeized graviton in this space. The equivalence
gives a possibility to calculate the intercept of the BFKL Pomeron
at large coupling constants~\cite{KLOV1, BPST}. The M\"{o}bius
invariance of the BFKL kernel was demonstrated also in two
loops~\cite{Moebtwo}. For next-to-leading calculations one can use
the effective field theory for reggeized gluons~\cite{eff}. The
generalized bootstrap equation gives a possibility to prove the
multi-Regge form of production amplitudes in the next-to-leading
approximation~\cite{Boot}.

Another application of the BFKL approach is a verification of the
BDS ansatz~\cite{BDS} for the inelastic amplitudes in
$N=4$ SUSY. It was demonstrated~\cite{BLS1, BLS2}, that the BDS amplitude $M^{BDS}_{2\rightarrow 4}$
should be
multiplied by the factor
containing the contribution of the Mandelstam cuts \cite{Mand} in LLA. In the two-loop approximation
this factor can be found
also from properties of analyticity
and factorization~\cite{Lip10}
or directly from recently obtained exact result \cite{Spradlin} for $M_{2\rightarrow
4}$ (see~\cite{LipPryg}). In a general case the wave function in LLA for the composite $n$-gluon state
in the adjoint representation satisfies
the Schr\"{o}dinger
equation for an open integrable Heisenberg spin chain~\cite{Intopen}.

In this paper we shall calculate the eigenvalues $\omega (t)$ of the kernel $K$
for the BFKL equation in the adjoint representation of the gauge group
at $N=4$ SUSY in the next-to-leading
approximation.
The Green function of this equation allows one to find the
asymptotic behavior of the inelastic amplitude in the Regge kinematics.
There is a hypothesis~\cite{AldMald, DHKS}, that the inelastic amplitude
with the maximal helicity violation in a planar  approximation is
factorized in the product of the BDS amplitude $M^{BDS}$, containing in
crossing channels the Regge factors with corresponding infrared
divergencies,
and the remainder function $R$ depending on the anharmonic ratios
\be
M=R\,M^{BDS}\,.
\ee
In an accordance with this hypothesis the $q^2$-dependence of the
eigenvalues of the octet BFKL equation is given by
the expression (cf. \cite{BLS2} in LLA)
\be
\omega (-q^2) =\omega _{g}(-q^2)+\omega _0\,,
\label{trajectory}
\ee
where $\omega _{g}(t)$ is the gluon Regge trajectory, which can be
expressed in all orders of the perturbation theory of $N=4$ SUSY
in terms of two functions entering in the expression for the
BDS amplitude~\cite{BLS1}. The "intercept" $\omega _0$ does not depend on
$q^2$ due  to the conformal invariance of $N=4$ SUSY and can be
written in terms of the "energy" $E=-\omega_0$ being the eigenvalue
of the BFKL kernel discussed in the next section.

\section{Integral kernel in the adjoint representation}

The homogeneous BFKL equation  can be
written in the form
\be
\omega _0 \phi = \hat{K}\phi \,,
\ee
where $\hat{K}$ is the integral operator from which the gluon Regge trajectory
is subtracted. In the momentum representation
it has the form
\be
\hat{K}\phi ( \vec q_1, \vec q_2)=\int
\frac{d^2q _1^{\;\prime}}{| q_1^{\;\prime}|^2| q_2^{\;\prime}|^2}
\,K(\vec q_1, \vec q_1^{\;\prime}; \vec q)\,\phi
(\vec q_1^{\;\prime}, \vec q_2^{\;\prime})\,,\,\,\vec q=
\vec q_1+\vec q_2= \vec q^{\;\prime}_1+\vec q^{\;\prime}_2\,.
\ee
The integral kernel for $N=4$ SUSY
can be presented as follows (cf.~\cite{BFKL, Boot} for the QCD case)
\begin{equation}
K(\vec q_1, \vec q_1^{\;\prime}; \vec q)= \delta^2(\vec q_1-\vec q_1^{\;\prime})
\vec q_1^{\;2}\vec q_2^{\;2}\left(\omega _{g}(-\vec q_1^{\;2})+\omega_g(-\vec q_2^{\;2})-
\omega_g(-\vec q^{\;2})\right) +
K_r(\vec q_1,\vec q_1^{\;\prime}; \vec q)\,, \label{octkernel}
\end{equation}
where the first term corresponds to virtual corrections with the
gluon regge trajectory subtraction (see (\ref{trajectory})) and the
second term appears from the real intermediate states in the
$s$-channel. The total contribution does not contain infrared
divergencies. Using results of Refs. \cite{Fadin:2007de} it can be written in the form
\begin{equation}
K(\vec q_1, \vec q_1^{\;\prime}; \vec q)= \frac{1}{2}\delta^2(\vec q_1-\vec q_1^{\;\prime})
\vec q_1^{\;2}\vec q_2^{\;2}\left(\omega _g(-\vec q_1^{\;2})+\omega_g(-\vec q_2^{\;2})-
2\omega_g(-\vec q^{\;2})\right) +
K^{ns}(\vec q_1,\vec q_1^{\;\prime}; \vec q),
\end{equation}
where
\be
\omega_g(-\vec q_1^{\;2})+\omega_g(-\vec q_2^{\;2})-
2\omega_g(-\vec q^{\;2})= - \frac{\alpha \,N_{c}}{2\pi}
\left(1-\zeta (2)\, \frac{\alpha \,N_{c}}{2\pi}
\right)\,\ln\left(
\frac{\vec{q}_1^{\;2}\vec{q}_2^{\;2}}{\vec{q}^{\;4}}\right)
\ee
and
\[
K^{ns}(\vec q_1, \vec q_1^{\;\prime}; \vec q)=-\delta^2(\vec{q}_{1}-\vec{q}^{\;\prime}_{1})
\vec{q}_1^{\;2}\vec{q}_2^{\;2}
\frac{\alpha \,N_{c}}{8\pi^{2}}\Biggl( \left(1-\zeta (2)\,\frac{\alpha \,N_{c}}{2\pi} \right)\int
d^{2}k\;\left(
\frac{2}{\vec{k}^{\;2}}+2\frac{\vec{k}
(\vec{q}_{1}-\vec{k})}{\vec{k}^{\;2}(\vec{q}_{1}-\vec{k})^2}\right)
\]
\[
-3\alpha \, N_{c}\zeta(3)
 \Biggr)
+\frac{\alpha \,N_{c}}{8\pi^{2}}\left\{
\Biggl(1-\zeta (2)\,\frac{\alpha \,N_{c}}{2\pi}
\Biggr)
\left(  \frac{\vec{q}_{1}^{\;2}\vec
{q}_{2}^{\;\prime\;2}+\vec{q}_{1}^{\;\prime\;2}\vec{q}_{2}^{\;2}}{\vec
{k}^{\;2}}-\vec{q}^{\;2}\right)+\right.
\]
\[
\frac{\alpha \,N_{c}}{4\pi}\Biggl[\frac{\vec{q}^{\,2}}{2}
\left(\ln\left(  \frac{\vec{q}_{1}^{\;2}}{\vec
{q}^{\;2}}\right)  \ln\left(  \frac{\vec{q}_{2}^{\;2}}{\vec{q}^{\;2}}\right)
+\ln\left(  \frac{\vec{q}_{1}^{\;\prime\;2}}{\vec{q}^{\;2}}\right)
\ln\left(  \frac{\vec{q}_{2}^{\;\prime2}}{\vec{q}^{\;2}}\right) +\ln^{2}\left(
\frac{\vec{q}_{1}^{\;2}}{\vec{q}
_{1}^{\;\prime\;2}}\right)  \right)
-\frac{\vec{q}_{1}^{\;2}\vec{q}
_{2}^{\;\prime\;2}
+\vec{q}_{2}^{\;2}\vec{q}_{1}^{\;\prime\;2}}{\vec{k}^{\;2}
}\ln^{2}\left(  \frac{\vec{q}_{1}^{\;2}}{\vec{q}_{1}^{\;\prime\;2}}\right)
\]
\[
 -\frac{1}{2}\,\frac{\vec{q}_{1}^{\;2}\vec{q}_{2}^{\;\prime\;2}-\vec{q}_{2}^{\;2}\vec{q}
_{1}^{\;\prime\;2}}{\vec{k}^{\;2}}\ln\left(  \frac{\vec{q}_{1}^{\;2}}{\vec
{q}_{1}^{\;\prime\;2}}\right)  \,\ln\left(
\frac{\vec{q}_{1}^{\;2}\vec{q}_{1}^{\;\prime\;2}}{\vec{k}^{\;4}}
\right) +\biggl[\vec{q}^{\;2}(\vec{k}^{\;2}-\vec{q}_{1}^{\;2}-\vec{q}
_{1}^{\;\prime\;2})
\]
\[
\left.+2\vec{q}_{1}^{\;2}\vec{q}_{1}^{\;\prime\;2}-\vec{q}
_{1}^{\;2}\vec{q}_{2}^{\;\prime\;2}-\vec{q}_{2}^{\;2}\vec{q}_{1}^{\;\prime
\;2}+\frac{\vec{q}_{1}^{\;2}\vec{q}_{2}^{\;\prime\;2}-\vec{q}_{2}^{\;2}\vec
{q}_{1}^{\;\prime\;2}}{\vec{k}^{\;2}}(\vec{q}_{1}^{\;2}-\vec{q}_{1}
^{\;\prime\;2})\biggr]I(\vec{q}_{1}^{\;2},\vec{q}_{1}^{\;\prime\;2},\vec
{k}^{\;2})\Biggr]\!\!\right\}
\]
\begin{equation}
+\left(  \vec{q}_{1}\leftrightarrow\vec{q}_{2},\;\;\vec{q}_{1}^{\;\prime
}\leftrightarrow\vec{q}_{2}^{\;\prime}\right)  ~, \label{ns kernel at D=4}
\end{equation}
where $\vec{k}=\vec{q}_{1}-\vec{q}_{1}^{\;\prime}$ and the function $I$
is given below
\begin{equation}
I(\vec{q}_{1}^{\;2},\vec{q}_{1}^{\;\prime\;2},\vec{k}^{\;2})=\int_{0}^{1}
\frac{dx}{\vec{q}_{1}^{\;2}(1-x)+\vec{q}_{1}^{\;\prime\;2}x-\vec{k}
^{\;2}x(1-x)}\ln\left(  \frac{\vec{q}_{1}^{\;2}(1-x)+\vec{q}_{1}^{\;\prime
\;2}x}{\vec{k}^{\;2}x(1-x)}\right)  ~. \label{integral I}
\end{equation}
Note that $I(a,b,c)$ is a totally symmetric function of the variables $a,\;b$
and $c$.

One could expect, that the BFKL kernel in $N=4$ SUSY is M\"{o}bius invariant in the
momentum representation, which would lead to the following simple form of its
eigenfunctions (cf. \cite{BLS2})
\be
\phi _{\nu n}( \vec q_1, \vec q_2)=\left|\frac{q_1}{q_2}\right|^{2i\nu}\,e^{i \,n\, \phi}\,,
\ee
where $\phi $ is the azimuthal angle of the complex number constructed
from transverse  components of the vectors $\vec{q}_{1}$ and $\vec{q}_{2}$
\be
\frac{q_1}{q_2}=\left|\frac{q_1}{q_2}\right|\,e^{i\phi}\,.
\ee

However, in the existing form the kernel is not M\"{o}bius invariant and
in future one should construct the similarity transformation to the  invariant
form (cf. \cite{Moebtwo}). Such transformation exists because the remainder function $R$,
corresponding to the correction factor for the BDS expression, should be
invariant under four-dimensional dual conformal transformations and
the Green function obtained from the BFKL equation in the adjoint
representation allows to find the asymptotic behavior of the remainder
function in the Mandelstam kinematical regions~\cite{BLS2}.

\section{Eigenvalues of the kernel}

It is important, that the eigenvalues
of the BFKL kernel do not depend on its representation and can be found
from our expression (\ref{octkernel}).
To calculate these eigenvalues we consider the BFKL equation in the limit (cf.~\cite{BLS2})
\be
|q_1|\sim |q'_1|\ll |q|\approx |q_2|\approx |q'_2|\,.
\ee
Denoting the two dimensional vectors $\vec q_1$ and $\vec q^{\;\prime}_1$ by
$\vec p$ and $\vec p^{\;\prime}$, respectively, we write the BFKL equation
in the form
\be
\int \frac{d^2p^{\;\prime}}{|p^{\;\prime}|^2} \,K(\vec p,\vec p^{\;\prime})\,
\Phi (\vec p^{\;\prime})=\omega_0 \,\Phi (\vec p)\,.
\ee
 Its kernel
 is given below
\[
K(\vec p,\vec p^{\;\prime})=-\delta^2(\vec p-\vec p^{\;\prime})\,|p|^2\,\frac{\alpha N_c}{4\pi ^2}\,
\left(\left(1- \frac{\alpha N_c}{2\pi }\zeta (2)\right)\,
\int d^2p^{\;\prime}\,\left(\frac{2}{|p^{\;\prime}|^2}+
\frac{2(p^{\;\prime},p-p^{\;\prime})}{|p^{\;\prime}|^2|p-p^{\;\prime}|^2}\right)-3\alpha \,\zeta (3)\right)
\]
\be
+\frac{\alpha N_c}{4\pi ^2}\,
\left(1- \frac{\alpha N_c}{2\pi }\zeta (2)\right)\,
\left(\frac{|p|^2+|p^{\;\prime}|^2}{|p-p^{\;\prime}|^2}-1\right)+
\frac{\alpha ^2N_c^2}{32\,\pi ^3 }\,R(\vec p,\vec p^{\;\prime})\,.
\ee
Here $\vec p$ and $\vec p^{\;\prime}$ are momenta of the same reggeized gluon before
and after its scattering in the $t_2$-channel (momenta of another gluon tend to infinity
together with $q$). The reduced kernel $R(\vec p,\vec p^{\;\prime})$ is given below
\be
R(\vec p, \vec p^{\;\prime})=\left(\frac{1}{2}-\frac{|p|^2+|p^{\;\prime}|^2}{|p-p^{\;\prime}|^2}\right)\,
\ln ^2\frac{|p|^2}{|p^{\;\prime}|^2}-\frac{|p|^2-|p^{\;\prime}|^2}{2|p-p^{\;\prime}|^2}\,
\ln \frac{|p|^2}{|p^{\;\prime}|^2}\,\ln \frac{|p|^2|p^{\;\prime}|^2}{|p-p^{\;\prime}|^4}
\ee
\be
+\left(-|p+p^{\;\prime}|^2+\frac{(|p|^2-|p^{\;\prime}|^2)^2}{|p-p^{\;\prime}|^2}\right)\,\int_0^1dx\,
\frac{1}{|(1-x)p+xp^{\;\prime}|^2}\,\ln \frac{|(1-x)p+xp^{\;\prime}|^2}{x(1-x)|p-p^{\;\prime}|^2}\,.
\ee
From the rotational and dilatational invariance of the kernel we obtain
its eigenfunctions in the simple form
\be
\Phi _{\nu n}(\vec p)=|p|^{2i\nu} e^{i\phi n}\,,
\ee
where $\phi$ is the angle of the transverse vector $\overrightarrow{p}$ with respect to
the axis $x$. Note, that $\nu $ is real and $n$ is integer.

The orthonormality condition for this set of functions is obvious
\be
\frac{1}{2\pi ^2}\int \frac{d^2p}{|p|^2}\,\Phi ^*_{\mu m}(\vec p)\,\Phi _{\nu n}(\vec p^{\;\prime})
=\delta (\mu -\nu )\,\delta _{m,n}\,.
\ee

The corresponding eigenvalues can be calculated with the action of the
BFKL kernel on the eigenfunctions and are given below
\be
\omega (\nu , n)=- a \left(E_{\nu n}+a\,\epsilon _{\nu n}\right)\,,\,\,
a=\frac{\alpha N_c}{2\pi }\,,
\label{omeganu}
\ee
where $E_{\nu n}$ is the "energy" in the leading approximation~\cite{BLS2}
\be
E_{\nu n}=-\frac{1}{2}\,\frac{|n|}{\nu ^2+\frac{n^2}{4}}+
\psi (1+i\nu +\frac{|n|}{2})+\psi (1-i\nu +\frac{|n|}{2})
-2\psi (1)\,,\,\,\psi (x)=(\ln \Gamma (x))'
\ee
and the next-to-leading correction $\epsilon _{\nu n}$ can be written as follows
\[
\epsilon _{\nu n}=-\frac{1}{4}\left(\psi ^{\prime \prime}(1+i\nu +\frac{|n|}{2})+
\psi ^{\prime \prime}(1-i\nu +\frac{|n|}{2})+\frac{2i\nu \left(\psi '(1-i\nu +\frac{|n|}{2})-\psi '(1+i\nu
+\frac{|n|}{2})\right)}{\nu ^2+\frac{n^2}{4}}
\right)
\]
\be
-\zeta (2)\,E_{\nu n}-3\zeta (3)-\frac{1}{4}\,\frac{|n|\,\left(\nu
^2-\frac{n^2}{4}\right)}{\left(\nu
^2+\frac{n^2}{4}\right)^3}\,.
\label{eigennext}
\ee
Here the $\zeta$-functions are expressed in terms of polylogarithms
\be
Li_n(x)=\sum _{k=1}^\infty \frac{x^k}{k^n}\,,\,\,\zeta (n)=Li_n(1)\,.
\ee

Note, that $\omega (\nu, n)$ has the important property
\be
\omega (0,0)=0\,.
\ee
It is in an agreement with the existence of the eigenfunction
$\Phi =1$ with a vanishing eigenvalue, which is a consequence
of the bootstrap relation~\cite{BFKL, Boot}.

\section{Corrections to the remainder function}

One can easily construct the Green function for the conformally invariant
BFKL kernel in terms of its eigenvalues. This Green function allows us
to calculate the remainder functions $R_n$ for an arbitrary number of external legs
in the regions, where there are Mandelstam's cuts corresponding
to the composite states of two reggeized gluons. For simplicity we consider the
remainder function $R_6$ for the gluon transition $2\rightarrow 4$ depending
on three anharmonic ratios (cf.~\cite{LipPryg})
\be
u_1=\frac{ss_2}{s_{012}s_{123}}\,,\,\,u_2=\frac{s_1t_3}{s_{012}t_2}\,,\,\,
u_3=\frac{s_3t_1}{s_{123}t_2}\,.
\ee
In the multi-regge kinematics  one obtains
\be
s\gg s_{012},s_{123}\gg s_1,s_2,s_3\gg t_1,t_2,t_3\,,
\ee
which corresponds to the following restrictions on the variables $u_k$
\be
1-u_1 \rightarrow 0\,,\,\,\tilde{u}_2=\frac{u_2}{1-u_1}\sim 1\,,\,\,\tilde{u}_3=
\frac{u_3}{1-u_1}\sim 1\,.
\ee
It is convenient also to introduce the complex variable $w$~\cite{LipPryg}
\be
w=|w|e^{i\phi _{23}}\,,\,\,|w|^2=\frac{u_2}{u_3}\,,\,\,\cos \phi _{23}=
\frac{1-u_1-u_2-u_3}{2\sqrt{u_2u_3}}
\ee
expressed in terms of transverse momenta of produced particles $k_1$, $k_2$
and momentum transfers $q_1,q_2,q_3$
\be
w=\frac{q_3k_1}{k_2q_1}\,.
\ee

In this case the remainder function $R$ in the Mandelstam region, where
\be
s,s_2\rightarrow +\infty\,,\,\,s_1,s_3\rightarrow -\infty \,,
\ee
can be presented in the form of a dispersion-like relation~\cite{Lip10}
\be
R\,e^{i\pi \delta}=cos \,\pi \omega _{ab}+i \,\frac{a}{2}
\sum _{n=-\infty}^\infty
(-1)^n\left(\frac{w}{w^*}\right)^{\frac{n}{2}}\int _{-\infty}^\infty\frac{|w|^{2i\nu}
d\nu}{\nu ^2 +\frac{n^2}{4}}
\,\Phi _{Reg} (\nu , n)\left(-\frac{1}{\sqrt{u_2u_3}}\right)^{\omega (\nu , n)},
\label{dispers}
\ee
where
\be
\delta =\frac{\gamma _K}{8}\,\ln (\tilde{u}_2\tilde{u}_3)=\frac{\gamma_K}{8}\,
\ln \frac{|w|^2}{|1+w|^4}\,,\,\,\omega _{ab}
=\frac{\gamma _K}{8}\,\ln \frac{\tilde {u}_2}{\tilde{u}_3}=\frac{\gamma _K}{8}\,\ln |w|^2
\ee
and the cusp anomalous dimensions
\be
\gamma _K =4a -4\,a^2\,\zeta (2)+22\,\zeta (4)\,a^3+...
\label{gammaK}
\ee
is known in
all orders of perturbation theory~\cite{BES}.

Further,  instead of the traditional variable $1/(1-u_1)$ (see~\cite{BLS1, BLS2})
we used in eq. (\ref{dispers})
the following
energy invariant
\be
\frac{1}{\sqrt{u_2u_3}}=s_2\,\frac{|q_2|^2}{\sqrt{|k_1|^2|q_1|^2}\,|k_2|^2|q_3|^2}=
\frac{1}{1-u_1}\,\frac{|1+w|^2}{|w|}\,,
\ee
because according to the Regge theory the amplitude should be factorized in the $t_2$-channel.
As a result, by expanding this expression for $R$ in the perturbation theory
\[
R=1+i\,a^2\left(b_1 \ln \frac{1}{1-u_1}+b_2\right)
+
a^3\left(ic_1\ln ^2\frac{1}{1-u_1}+(d_1+ic_2)\ln\frac{1}{1-u_1}+d_2+ic_3\right)+...=
\]
\be
1+i\,a^2\left(\widetilde{b}_1 \ln
\frac{1}{\sqrt{u_2u_3}}+\widetilde{b}_2\right)
+
a^3\left(i\widetilde{c}_1\ln ^2\frac{1}{\sqrt{u_2u_3}}+(\widetilde{d}_1+i\widetilde{c}_2)
\ln\frac{1}{\sqrt{u_2u_3}}+\widetilde{d}_2+i\widetilde{c_3}\right)+...\,,
\label{expand}
\ee
we obtain~\cite{BLS2, LipPryg}
\be
\widetilde{b}_1=b_1=-\frac{\pi}{2}\,\ln |1+w|^2\ln\frac{|1+w|^2}{|w|^2}\,,
\ee
\[
\widetilde{b_2}=b_2-b_1\ln\frac{|1+w|^2}{|w|}\,,\,\,
\frac{1}{\pi} \,
b_2=\frac{1}{2}\ln |w|^2\ln ^2|1+w|^2
\]
\be
-\frac{1}{3}\ln^3|1+w|^2+\ln |w|^2
\left(Li_2 (-w)+Li_2(-w^*)\right)-2\left(Li_3(-w)+L_3(-w^*)\right)\,,
\label{b1b2}
\ee
and (see ref.~\cite{LipPryg})
\[
\frac{4}{\pi }\,\widetilde{c}_1=\frac{4}{\pi }\,
c_1=\ln |w|^2\,\ln ^2|1+w|^2-\frac{2}{3}\ln^3|1+w|^2-\frac{1}{4}\ln ^2|w|^2
\ln |1+w|^2
\]
\be
+\frac{1}{2}\,\ln |w|^2\left(Li_2(-w)+Li_2(-w^*)\right)-Li_3(-w)-Li_3(-w^*)\,,
\ee
\[
\frac{4}{\pi ^2}\,\widetilde{d}_1
=\frac{4}{\pi ^2}\,d_1=-\ln |w|^2\,\ln ^2|1+w|^2+\frac{2}{3}\ln^3|1+w|^2+\frac{1}{2}\ln ^2|w|^2
\ln |1+w|^2
\]
\be
+\ln |w|^2\left(Li_2(-w)+Li_2(-w^*)\right)-2Li_3(-w)-2Li_3(-w^*)\,.
\ee
Note, that in the second order the real contribution to $R$ is absent~\cite{Lip10}.

The product of two impact factors $\Phi _{Reg}(\nu, n)$ can be obtained with the use of the
Fourier transformation of the function $\widetilde{b}_2$
\be
\Phi _{Reg}(\nu, n)=1+\Phi _{Reg}^{(1)}(\nu, n)\,a+\Phi _{Reg}^{(2)}(\nu, n)\,a^2+...\,,
\label{impactexp}
\ee
\be
\Phi _{Reg}^{(1)}(\nu, n)=\Phi ^{(1)}(\nu , n)+\Delta \Phi (\nu ,n) =
-\frac{1}{2}E_{\nu n}^2-\frac{3}{8}\,\frac{n^2}{\left(\nu ^2+\frac{n^2}{4}
\right)^2}-\zeta (2)\,,
\label{impact}
\ee
where $\Delta \Phi (\nu ,n)$ is the contribution of the term $-b_1\ln\frac{|1+w|^2}{|w|}$
in $\widetilde{b}_2$ (\ref{b1b2}) and
the contribution $\Phi ^{(1)}(\nu , n)$ appearing from the term $b_2$
was calculated in ref.~\cite{LipPryg}
\footnote{In the reference~\cite{LipPryg}
the
quantity $\Phi ^{(1)}(\nu, n)$ was found for the remainder function, but here we need it
for the full amplitude. According to (\ref{dispers}) they differ by the term appearing
from the expansion of $\exp (i\pi \delta)$ and proportional to
the second order contribution to the anomalous dimension $\gamma _K$ (\ref{gammaK}).}
\be
\Phi ^{(1)}(\nu, n)=
E_{\nu n}^2-\frac{1}{4}\,\frac{n^2}{\left(\nu ^2+\frac{n^2}{4}
\right)^2}-\zeta (2)\,.
\ee

The knowledge of eigenvalues (\ref{eigennext})
in the next-to-leading approximation gives a possibility to calculate
the coefficients $\widetilde{c}_2$ and $\widetilde{d}_2$
from expression (\ref{dispers})
\[
\frac{1}{\pi}\,\widetilde{c}_2=-\frac{1}{4}\,\ln |w|^2 \left(S_{1,2}(-w)+S_{1,2}(-w^*)
+\ln (1+w)\,Li_2(-w)+\ln (1+w^*)\,Li_2(-w*)\right)
\]
\[
+\frac{\zeta (3)}{2}\,\ln |1+w|^2-
\ln \frac{|1+w|^2}{|w|} \left(Li_3 (-w)+Li_3 (-w^*)-\frac{1}{2}\ln
|w|^2
(Li_2 (-w)+Li_2 (-w^*))\right)
\]
\[
+\frac{1}{4}\,\ln |1+w|^2(Li_3 (-w)+Li_3 (-w^*))+\frac{1}{16}\,\ln ^2|w|^2\ln |1+w|^2\ln \frac{|1+w|^2}{|w|^2}
\]
\be
+\frac{1}{8}\,\ln^2 |1+w|^2
\ln^2 \frac{|1+w|^2}{|w|^2}
+\frac{1}{8}\ln ^2|w|^2\ln (1+w)\,\ln (1+w^*)+\zeta (2)\,
\ln |1+w|^2\ln \frac{|1+w|^2}{|w|^2}\,,
\label{c2}
\ee
\be
\widetilde{d}_2=\pi \left(\widetilde{c}_2-\ln \frac{|1+w|^2}{|w|}
\,\widetilde{b}_2+2\,\zeta (2)\,\widetilde{b}_1\right)\,.
\label{d2}
\ee
In the above expression
the function $S_{1,2}(-x)$ has the following representation
\be
S_{1,2}(-x)=\int _{0}^x\frac{dx'}{2x'}\,\ln^2 (1+x')=Li _3(\frac{x}{1+x})+Li_3(-x)-
\ln (1+x)Li _2(-x)-\frac{1}{6}\ln ^3(1+x)\,.
\ee
One can verify with the use of the known relations among polylogarithms $Li_n (x)$,
that the coefficients $\widetilde{c}_2$ and $\widetilde{d}_2$ are single-valued functions
on the two-dimensional plane $\overrightarrow{w}$ and are symmetric to the inversion
$w\rightarrow 1/w$.
We can calculate also the coefficients $c_2$ and $d_2$ in (\ref{expand}) using the relations
\be
c_2=\widetilde{c}_2+2\widetilde{c}_1\ln \frac{|1+w|^2}{|w|}\,,\,\,
d_2=\widetilde{d}_2+\widetilde{d}_1\,\ln \frac{|1+w|^2}{|w|}\,.
\label{c2d2}
\ee

Note, that recently the authors of  ref.~\cite{DDH} suggested an anzatz for the remainder function
$R_6$ in three loops based on the theory of symbols. They calculated its high energy behavior
in our Mandelstam region in the form of the polynomial expansion in $\log (1-u_1)$.
It turns out, that up to three loops their results are completely coincides with our
perturbative expansion (\ref{expand}). In particular, one can derive the expressions
(58) and (66) from the paper~\cite{DDH}  using the fact, that
the corresponding functions $g_1^{(2)}(w,w^*)$ and $h_0^{(3)}(w,w^*)$ are related with our coefficients
$c_2$ and $d_2$ in (\ref{expand}) as follows
\be
g_1^{(2)}(w,w^*)=-\frac{c_2}{2\pi}\,,\,\,h_0^{(3)}(w,w^*)=-\frac{d_2}{(2\pi )^2}\,.
\ee
It gives a possibility to fix the parameters $\gamma '$ and $\gamma '''$ appearing in
ref.~\cite{DDH} in the form
\be
\gamma '=-\frac{9}{2}\,,\,\,\gamma'''=0\,.
\ee
In expression (63) of the paper~\cite{DDH} also the additional
function $g_0^{(3)}(w,w^*)$ was calculated.
This function contains three unknown parameters appearing in the last line of (63). Our coefficients
$c_3$ and $\widetilde{c}_3$ in (\ref{expand}) can be expressed in terms of it
\be
c_3=2\pi \,g_0^{(3)}(w,w^*)\,,\,\,\widetilde{c}_3=c_3-\ln \frac{|1+w|^2}{|w|}\,c_2+
\ln ^2\frac{|1+w|^2}{|w|}\,c_1\,.
\ee
It gives a possibility to construct the following function
\be
\rho (w, w^*)=\frac{\widetilde{c}_3}{\pi} +\pi \,\widetilde{c}_1+\ln \frac{|1+w|^2}{|w|}\,
\left(\zeta (2)\,\ln ^2 \frac{|1+w|^2}{|w|}-\frac{11}{2}\,\zeta (4)\right),
\ee
where the term proportional to $\zeta (4)$ appears from the third order contribution
to $\gamma _K$ (\ref{gammaK}) which was calculated firstly in ref.~\cite{KLOV}. The
important next-to-next-to-leading
corrections to the product of impact-factors $\Phi _{Reg}(\nu, n)$ (\ref{impactexp})
can be expressed through $\rho (w, w^*)$
\be
\Phi ^{(2)}_{Reg}(\nu, n)=(-1)^n\left(\nu ^2+\frac{n^2}{4}\right)\int \frac{d^2w}{\pi}\,
\rho (w, w^*)\,|w|^{-2i\nu -2}\,\left(\frac{w^*}{w}\right)^{\frac{n}{2}}\,.
\ee
We are going to calculate $\Phi ^{(2)}_{Reg}(\nu, n)$ in future.

Similar results can be obtained for the remainder function describing the $3\rightarrow 3$
transitions in the corresponding Mandelstam regions~\cite{threethree}.

\section{Collinear limit}

It is well known, that the BFKL equation for the Pomeron wave function gives a possibility to
predict the leading singularities of the anomalous dimensions $\gamma$ of the twist-2 operators
at $\omega \rightarrow 0$ in
all orders of perturbation theory~\cite{BFKL, FL}. In particular, for the case of $N=4$ SUSY
the predictions of Ref.~\cite{trajN4} are in a full agreement with the direct calculations
of $\gamma$ up to 5 loops~\cite{KLOV, KLRSV, LRV}. As it follows from
the previous section, the BFKL kernel for the adjoint representation of the gauge group
allows one to find the high energy corrections to the remainder functions. On the other hand,
in the collinear limit the remainder functions obey the renormalization group-like
equations~\cite{AGMSV, GMSV}. The analytic continuation of the collinear expressions for $R$ to the
Mandelstam regions was performed in Ref.~\cite{BLP}. The leading asymptotics corresponds
to the unit conformal spin $|n|=1$. The anomalous dimensions $\gamma _{col}$ for the collinear limit
in the Euclidean region were
constructed~\cite{GMSV} and the relation between the Regge and collinear limits was
investigated~\cite{BLP}.

To calculate $\gamma _{col}$ in the Mandelstam region we present expression (\ref{dispers})
in the following form with the use of the Fourier transformation
\be
R\,e^{i\pi \delta}=cos \,\pi \omega _{ab}+i \,\frac{a}{2}
\sum _{n=-\infty}^\infty
(-1)^n\left(\frac{w}{w^*}\right)^{\frac{n}{2}}\int _{-\infty}^\infty d\nu \,|w|^{2i\nu}
\,L_{\nu n}\left(-\frac{1}{1-u_1}\right),
\label{dispers1}
\ee
where
\be
L _{\nu n}\left(-\frac{1}{1-u_1}\right)=\sum _{n'=-\infty}^\infty
(-1)^{n'-n}\int _{-\infty}^\infty\frac{\Phi _{reg}(\nu' ,n')
d\nu '}{\nu ^{\prime 2} +\frac{n^{\prime 2}}{4}}\,S_{\nu 'n'}^{\nu n}\,
\left(-\frac{1}{1-u_1}\right)^{\omega (\nu' ,n')}
\label{Lnun}
\ee
and
\be
S_{\nu 'n'}^{\nu n}=\int \frac{d^2w}{2\pi ^2}\,|w|^{2i(\nu'-\nu)-2}\,
\left(\frac{w}{w^*}\right)^{\frac{n'-n}{2}}\,\left(\frac{|1+w|^2}{|w|}\right)^{\omega (\nu' n')}\,.
\ee
The collinear limit $w\rightarrow 0$ or
$w \rightarrow \infty$ of the remainder function (\ref{dispers1}) should be performed
at fixed $1-u_1$~\cite{GMSV, BLP}.
Generally expressions (\ref{dispers1}) and (\ref{Lnun}) correspond to the collinear
renormalization with an infinite number of the multiplicatively renormalizable operators
(cf.~\cite{BLP}). But in the case, when we take into account only the asymptotic terms
at $|w|\rightarrow \infty$ with the conformal spin $|n|=1$, we can obtain
for $R$ the simple expression
\be
R\,e^{i\pi \delta} \approx cos \,\pi \omega _{ab}-ia\,\cos \phi _{23}\,
|w|^{-1}\int _{-i\infty}^{i\infty} d\omega \,
\frac{\Phi ^{Reg}(\nu ,1 )}{\left(\nu ^2+\frac{1}{4}\right)\,
\frac{d\omega}{d \nu }}\,|w|^{2\gamma _{col}(\omega)}\,
\left(-\frac{1}{1-u_1}\right)^\omega \,,
\ee
where the contour of integration goes to the right
of the BFKL singularity $\nu \sim \sqrt{\omega -\omega (0,1)}$ present in the integrand
in an accordance with the fact, that
the functions $\gamma =\gamma _{col}(\omega),\,\nu =\nu (\omega
)$ satisfy the set of equations
\footnote{Note, that our definition of the collinear anomalous dimension $\gamma _{col}$
differs with the factor $-1/2$ from that used in ref.~\cite{GMSV}.}
\be
\gamma =\frac{1}{2}+i\nu +\frac{\omega }{2}\,,\,\,\omega =
\omega (\nu,1)\,.
\label{shift}
\ee
For finding $\gamma_{col}$ in perturbation theory the function $\omega (\nu , 1)$
(\ref{omeganu}) should be expanded near the point $\nu =i/2 $
\be
\lim _{\nu \rightarrow \frac{i}{2}}\omega (\nu, 1)= \frac{a}{2}\,f_1\left(i\nu +\frac{1}{2}\right)
-\frac{a^2}{8}\,f_2\left(i\nu +\frac{1}{2}\right)\,,
\ee
where
\be
f_1(x)=\frac{1}{x}-1-x-x^2(1-4\zeta (3))-x^3+O(x^4)\,,
\ee
\be
f_2(x)=\frac{1}{x ^3}+\frac{1}{x ^2}+\frac{4\zeta (2)}{x }-8 \zeta (3)-4\zeta (2)-2+O(x)
\,,
\ee
Thus, we obtain the following equation for $\gamma =\gamma _{col}(\omega)$
\be
\omega =\frac{a}{2}\,f_1(\gamma )-\frac{a^2}{8}\,\left(f'_1(\gamma )f_1(\gamma )+f_2 (\gamma )\right)\,.
\ee
Its perturbative solution is given below
\be
\gamma _{col}(\omega )=\frac{a}{2}\,\left(\frac{1}{\omega}-1\right)-\frac{a^2}{4}
\left(\frac{1}{\omega ^2}+2\,\frac{\zeta (2)}{\omega}\right)+\frac{a^3}{4\,\omega ^2}\left(1+2\zeta (2)
+\zeta (3)\right)+O(a^4)\,.
\label{gamcol}
\ee
The above approach is similar to that for the singlet BFKL kernel, but in that case one obtained
the main contribution
to the Bjorken limit from $n=0$~\cite{FL}.

The collinear anomalous dimension $\gamma _{col}(\omega )$ in the Mandelstam region $s,s_2>0,s_1,s_3<0$
can be found in one loop using the results of the paper~\cite{BLP}.
We start with the perturbative expansion of the remainder function in the collinear limit
$|w|\rightarrow \infty$ in LLA
of the Operator Product Expansion (OPE)~\cite{AGMSV}
\be
R_{OPE}\approx a\cos \phi\,\frac{e^{-\sigma}}{2|w|}\,\sum _{k=0}^\infty \frac{(-a\ln |w|)^k}{k!}\,h_k(\sigma )
\,,\,\,\sigma =\frac{1}{2}\,\ln \frac{u_1}{1-u_1}\,,
\ee
where we expressed the world sheet coordinates $\tau$ and $\sigma$ in terms of our variables $w$ and $u_1$ (see
eqs (76)-(79) from ref.~\cite{BLP})
and included one loop contribution contained in the BDS amplitude.
The analytic continuation of the two loop remainder function calculated in ref.~\cite{Spradlin} to the Mandelstam
region $s,s_2>0,\,s_1,s_3<0$ gives the result (see eqs. (51), (C.12)-(C.16) from ref.~\cite{BLP})
\[
\cos \phi\,\rightarrow \cos \phi _{23};\;\; h_k(\sigma)\rightarrow  -h_k(\sigma)+\Delta_k(\sigma)\,,\,\,\frac{\Delta_0(\sigma)}{2\pi i}=-2\,e^\sigma \,,\,\,
\]
\be
\frac{\Delta_1(\sigma)}{2\pi i}=4\left(\cosh \sigma \ln (1+e^{2\sigma})-e^\sigma  \right).
\label{h1h2}
\ee
Here the functions $h_k(\sigma)$ for $k=0,1$ in the right hand side of the first relation are known from
ref.~\cite{GMSV}. They are
not essential for the calculation of $\gamma_{col}$ because they are real and fall at large $\sigma$.
The contributions $\Delta _k(\sigma)$ appear from the analytic continuation of the
corresponding discontinuities of the functions $h_k(\sigma)$ on the cut
$-1<\widetilde{s}_2<0$, where $\widetilde{s}_2=\exp (2\sigma)$~\cite{BLP}. After the continuation we can
write this discontinuity using the collinear renormalization group in the form
\be
\Delta R_{OPE}=-\,a\,\cos \phi _{23}\, \frac{1}{|w|}\,\int _{-i\infty}^{i \infty}
\frac{d\omega}{\omega (\omega +1)}\,|w|^{2\gamma _{col}(\omega) }e^{2\omega\sigma}\,,
\ee
where
\[
\gamma _{col}(\omega )=a\,\omega \,(1+\omega )\,\int _0^\infty d(2\sigma )\, e^{-2\sigma \omega }
\left(e^{-\sigma }\cosh (\sigma )\ln\left(1+e^{2\sigma}\right)-1\right)=
\]
\be
\frac{a}{2}\left(\frac{1}{\omega (\omega +1)}-2\omega
+(\omega +1)\left(\psi (\omega +1)-\psi (\frac{\omega +2}{2})\right)+
\omega \,\psi (\omega +2)-\omega \, \psi
(\frac{\omega+3}{2})\right).
\ee

As one can see from expression (\ref{gamcol}), the BFKL approach reproduces
correctly the first two terms of $\gamma _{col}$ at $\omega \rightarrow 0$.

\section{Conclusion}
In this paper we solved the BFKL equation for the channel with color octet quantum numbers
in the next-to-leading
approximation. The eigenvalues of its integral kernel were used to calculate in the
next-to-leading logarithmic approximation the
remainder function for the production amplitude $2\rightarrow 4$ in the multi-Regge kinematics
at the Mandelstam channels. The obtained result in
three loops is in an agreement with the recently suggested anzatz~\cite{DDH}
for the remainder function.
This anzatz allowed us to construct the product of corresponding impact-factors in the
next-to-next-to-leading approximation.
The collinear anomalous dimension in the Mandelstam region was calculated
explicitly in one loop. Its leading and  next-to-leading
singularities are found in all loops.

\vspace{0.5cm} {\textbf{ Acknowledgments}.}
We thank J. Bartels, A. Prygarin and G.  Vacca for helpful discussions,
the Hamburg University and DESY for  the warm hospitality and support. This work was done
in the framework of the program LEXI "Connecting Particles with the Cosmos".

\end{document}